\documentclass[a4paper,11pt]{article}
\usepackage{siunitx}
\usepackage{pos}
\usepackage{calrsfs}
\usepackage{bm}
\usepackage{cleveref}

\crefname{section}{section}{sections}
\Crefname{section}{Section}{Sections}

\crefname{figure}{figure}{figures}
\Crefname{figure}{Figure}{Figures}
\crefrangeformat{figure}{figures~#3#1#4--#5#2#6}
\Crefrangeformat{figure}{Figures~#3#1#4--#5#2#6}

\crefrangeformat{equation}{eqs.~(#3#1#4)--(#5#2#6)}
\Crefrangeformat{equation}{Equations~(#3#1#4)--(#5#2#6)}

\creflabelformat{subequations}{(#2#1#3)}
\crefname{subequations}{eqs.}{eqs.}
\crefrangeformat{subequations}{eqs.~(#3#1#4)--(#5#2#6)}
\Crefname{subequations}{Equations}{Equations}
\Crefrangeformat{subequations}{Equations~(#3#1#4)--(#5#2#6)}

\newcommand{\as}{\ensuremath{\alpha_\text{s}}}

\newcommand{\an}[1]{\ensuremath{\alpha_{#1}}}
\newcommand{\dd}{\ensuremath{\mathrm{d}}}
\newcommand{\e}{\ensuremath{\mathrm{e}}}
\newcommand{\iunit}{\ensuremath{\mathrm{i}}}
\newcommand{\case}[2]{\ensuremath{{\textstyle\frac{#1}{#2}}}}
\newcommand{\half}{\ensuremath{\case{1}{2}}}

\newcommand{\quarter}{\ensuremath{\case{1}{4}}}

\newcommand{\LQCD}{\ensuremath{{\Lambda_\text{QCD}}}}

\newcommand{\MSbar}{\ensuremath{\overline{\rm MS}}}

\newlength{\figwidth}

\title{Factorial growth in perturbation theory, power corrections: precise extraction of quark masses and~$\as$}
 \ShortTitle{Factorial growth}

\author*{Andreas S. Kronfeld}

\affiliation{Theory Division, Fermi National Accelerator Laboratory, \\
  PO Box 500, Batavia, IL 60510, USA}
\affiliation{Institute for Advanced Study, Technische Universität München,\\
  Lichtenbergstraße 2\,a, 85748 Garching, Germany}

\emailAdd{ask@fnal.gov}

\abstract{These proceedings summarize a newly found connection between the factorial growth of coefficients in perturbative QCD and power 
corrections to the perturbation series, discussed in refs.~\cite{Kronfeld:2023jab, Komijani:2017vep, Brambilla:2017hcq, Kronfeld:2024qao}.
The improved convergence is shown for three quantities four which four terms in the series are available: the static energy, the 
quark pole mass, and the polarized Bjorken sum rule.
Prospects for determinations of \as\ with controlled truncation uncertainties are discussed, as was found earlier in quark-mass 
determinations~\cite{Brambilla:2017hcq, FermilabLattice:2018est}.}

\FullConference{The XVIth Quark Confinement and the Hadron Spectrum Conference (QCHSC24)\\
 19-24 August, 2024\\
 Cairns Convention Centre, Cairns, Queensland, Australia\\}

\begin{document}
\maketitle

\section{Introduction}
\label{sec:intro}

The strong coupling \as\ is determined by measuring (or computing nonperturbatively) a physical quantity dominated by a hard scale and analyzing it in the context of perturbation theory: which value \as\ reproduces the physical quantity at hand?
A general obstacle to success is the truncation of the perturbative series~\cite{ParticleDataGroup:2024cfk,FlavourLatticeAveragingGroupFLAG:2024oxs}, because the perturbative series does not converge.
Perturbation theory develops an asymptotic series, \emph{i.e.}, a series with vanishing radius of convergence.
These proceedings summarize a new yet simple and general way to address this problem~\cite{Kronfeld:2023jab, Komijani:2017vep, Brambilla:2017hcq, Kronfeld:2024qao}, with applications to the static energy, the pole mass, and the polarized Bjorken sum rule.

Consider a physical observable with a single hard scale~$Q$, and form a dimensionless combination of the observable and powers of~$Q$, denoted $\mathcal{R}(Q)$.
Factorization in perturbative QCD then says, for $Q\gg\Lambda$,
\begin{equation}
    \mathcal{R}(Q) = r_{-1} + R(Q) + C_p\frac{\Lambda^p}{Q^p},
    \label{eq:effcharge}
\end{equation}
where $\Lambda$ is the scale characteristic of QCD, $r_{-1}$ is the value in the absence of QCD interactions, and $R(Q)$ and 
$C_p\Lambda^p/Q^p$ are called the ``perturbation series'' and the ``power correction'', respectively.
The perturbation series is written as
\begin{equation}
    R(Q) = \sum_{l=0} r_l(\mu/Q) \as(\mu)^{l+1} .
    \label{eq:series}
\end{equation}
Here $\mu$ is the renormalization scale and also (at least roughly speaking) a separation scale: energy scales above (below) $\mu$ 
are supposed to be described by the perturbation series $R(Q)$ (the power correction $C_p\Lambda^p/Q^p$).
The right-hand side of \cref{eq:effcharge} can be justified in a variety of ways: an operator-product expansion, an effective field theory, study of loop integrands, phenomenological reasoning, etc.
In general, more than one power correction can appear in \cref{eq:effcharge}, which can also addressed by the ideas summarized in 
these proceedings, as discussed briefly below.

The separation into high and low energies (or short and long distances) could be implemented \`a la Wilson~\cite{Wilson:1974mb}, but particle physicists most often obtain the perturbative coefficients $r_l(\mu/Q)$ via dimensional regularization with modified minimal subtraction of ultraviolet divergences (\emph{i.e.}, the  \MSbar\ scheme).
Then $\mu$ does not physically separate scales, rendering the separation between short and long distances (mathematically) ambiguous.
In particular, the coefficients of the asymptotic series, the $r_l$ in \cref{eq:series}, grow factorially with $l$, in a way 
dictated by the power~$p$ in \cref{eq:effcharge}, and the ambiguity is of order $\e^{-p/2\beta_0\as}$.%
\footnote{The $\mu$~dependence of $\as(\mu)$ in \cref{eq:series} is dictated by the QCD beta function, 
$\dd\as/\dd\ln\mu=-2\beta_0\as^2-2\beta_1\as^2-\cdots$.}
The results given below resolve this ambiguity, and similar ambiguities from higher powers, in a specific way.

\section{Factorial growth: some background}
\label{sec:fgb}

Factorial growth in asymptotic series is ubiquitous, appearing in quantum mechanics~\cite{Bender:1971gu, Bender:1973rz}, simple 
field theories~\cite{Gross:1974jv}, and QED~\cite{Lautrup:1977hs}.
A pedagogical example is the integral~\cite{Duncan:2012aja}
\begin{equation}
    Z(\lambda/m^4) = \frac{m}{\sqrt{2\pi}} \int_{-\infty}^{\infty} \dd\phi\, \e^{- \half m^2\phi^2 - \quarter\lambda\phi^4} ,
\end{equation}
which admits an asymptotic expansion for small $\alpha\equiv\lambda/m^4$
\begin{equation}
    Z(\alpha) \approx 1 + \sum_{l=0} a_l \alpha^{l+1}, \quad a_l = (-1)^{l+1}
        \frac{\Gamma(2l+\case{5}{2})}{\sqrt{\pi}\Gamma(l+2)},
    \label{eq:tony}
\end{equation}
obtained by expanding out the exponential $\e^{-\quarter\lambda\phi^4}$ before integrating over $\phi$.
For large $\lambda$, on the other hand, expanding $\e^{-\half m^2\phi^2}$ before integrating leads to the convergent expansion
\begin{equation}
    Z = \alpha^{-1/4} \sum_{j=0}^\infty c_j \alpha^{-j/2}, \quad c_j = \frac{(-1)^j}{2\sqrt{\pi}} 
        \frac{\Gamma(\frac{j}{2}+\quarter)}{\Gamma(j+1)}.
    \label{eq:toy}
\end{equation}
Because \cref{eq:toy} is a convergent series, the sum reproduces the exact expression
\begin{equation}
    Z(\alpha) = \frac{\e^{1/8\alpha}}{2\sqrt{\pi\alpha}} K_{1/4}(1/8\alpha)
    \label{eq:eggzact}
\end{equation}
with a Bessel function~$K_\nu$.
A procedure known as Borel summation can be applied to \cref{eq:tony}: because the $a_l$ alternate in sign, the outcome is 
mathematically unambiguous and yields \cref{eq:eggzact} too.

In a similar vein, it has long been known (see, \emph{e.g.}, ref.~\cite{Beneke:1998ui}) that the QCD coefficients satisfy
\begin{equation}
    r_l \sim R_0^{(p)} \left(\frac{2\beta_0}{p}\right)^l \frac{\Gamma(l+1+pb)}{\Gamma(1+pb)} , 
    \label{eq:asymp-rl}
\end{equation}
where ``$\sim$'' means ``goes asymptotically as''---in other words, the relation is supposed to hold only for large orders $l\gg1$.
The arguments of the $\Gamma$ (\emph{i.e.}, factorial!) function contain the quantity $b=\beta_1/2\beta_0^2$, where $\beta_0$ and $\beta_1$ are the every-scheme coefficients of the beta function.
For $n_f=3$ flavors, $b=32/81\approx0.4$.
The $l$-independent factor $R_0^{(p)}$ in \cref{eq:asymp-rl} is the ``normalization'' or ``strength''.
Expressions for the strength exist in the literature~\cite{Lee:1996yk, Lee:1999ws, Pineda:2001zq, Hoang:2008yj}, but the 
derivations and/or explicit outcomes are hard (for me) to understand.

For the gluonic energy stored between a static quark and a static antiquark (the ``static energy'' for short), the first four 
coefficients ($l\in\{0,1,2,3\}$) are 1, 1.38, 5.46, and 26.7.
It looks as if the factorial growth appears already for accessible~$l$.

\section{Factorial growth in QCD: renormalization constraints}
\label{sec:fgrg}

Returning to \cref{eq:effcharge,eq:series}, the physical quantity $\mathcal{R}(Q)$ cannot depend on the artificial renormalization scale $\mu$ of the \MSbar~scheme.
The $\mu$ independence of $\mathcal{R}(Q)$ imposes constraints on the $\mu/Q$ dependence of the $r_l(\mu/Q)$, such that the $\mu$~dependence of $\as(\mu)$ is cancelled.
In this way, $r_l(\mu/Q)$ contains pieces of the form $r_{l-j}\beta_j\,(\ln\mu/Q)^j$, $1\le j\le l$.
Remarkably and crucially, the only source of $Q$ dependence (well, when all quarks can be taken massless or decoupled) of $R(Q)$ springs from the $(\ln\mu/Q)^j$ pieces in the $r_l(\mu/Q)$.

This renormalization constraints on the $Q$ dependence imply a connection between the perturbative series and the power correction.
An observable without the power correction can be obtained from $\mathcal{R}$ by defining~\cite{Kronfeld:2023jab, Komijani:2017vep, 
Brambilla:2017hcq, Kronfeld:2024qao}
\begin{equation}
    \mathcal{F}^{(p)}(Q) \equiv \frac{1}{pQ^{p-1}} \frac{\dd Q^p\mathcal{R}}{\dd Q} 
            = r_{-1} + F^{(p)}(Q) = r_{-1} + \sum_{k=0} f_k^{(p)}(\mu/Q) \as(\mu)^{k+1} ,
    \label{eq:FQ}
\end{equation}
where
\begin{equation}
    f^{(p)}_k = r_k - \frac{2}{p} \sum_{j=0}^{k-1} (j+1)\beta_{k-1-j}r_j.
    \label{eq:fk}
\end{equation}
\Cref{eq:fk} has the structure of a matrix equation $\bm{f}^{(p)}=\mathbf{Q}^{(p)}\cdot\bm{r}$ with a lower-diagonal matrix~$\mathbf{Q}^{(p)}$.
Even though the matrix and the vectors $\bm{r}$ and $\bm{f}^{(p)}$ are infinite, the equation can be solved for $\bm{r}$ row-by-row.
The solution is
\begin{equation}
    r_l = f_l^{(p)} + \left(\frac{2\beta_0}{p}\right)^l \frac{\Gamma(l+1 + pb)}{\Gamma(1 + pb)}
            \sum_{k=0}^{l-1} (k + 1) \frac{\Gamma(1 + pb)}{\Gamma(k+2 + pb)} \left(\frac{p}{2\beta_0}\right)^k f_k^{(p)} ,
    \label{eq:rl}
\end{equation}
This result shares with \cref{eq:asymp-rl} the same growth $(2\beta_0/p)^l\Gamma(l+1+pb)/\Gamma(1+pb)$, now with an explicit 
strength, which is the mildly $l$-dependent sum $\sum_{k=0}^{l-1}\cdots f_k^{(p)}$.
A slight difference is the extra term, $f_l^{(p)}$, and a major difference is that the relation is ``$=$'' instead of ``$\sim$''.
\Cref{eq:rl} holds at every order, hence already at low orders.

The coefficients $f_l^{(p)}$ can grow, just not as fast as the $r_l$.
The manipulations in \cref{eq:FQ,eq:fk,eq:rl} do not prove this property, but consistency with the methods~\cite{Beneke:1998ui} yielding \cref{eq:asymp-rl} require it.
For example, if there is a second power correction with $p'>p$, then the $f_l^{(p)}$ satisfy a formula similar to \cref{eq:rl} but with~$p'$ replacing~$p$.
This growth is slow enough for the sum defining the strength to converge, hence the assertion that the strength depends mildly on~$l$.

In practice, the first $L$ coefficients $r_l$ are available in the literature, and just as many $f^{(p)}_k$ are obtained from them via \cref{eq:fk}.
So how can \cref{eq:rl} yield anything new?
The answer to this question is that for $l<L$, \cref{eq:rl} regenerates the original $r_l$, as it must.
For $l\ge L$, however, \cref{eq:rl} provides a compelling approximation, namely
\begin{align}
    r_l \approx R_l^{(p)} &\equiv
        R_0^{(p)} \left(\frac{2\beta_0}{p}\right)^l \frac{\Gamma(l+1 + pb)}{\Gamma(1 + pb)} , \qquad l\ge L,
    \label{eq:approx-rl} \\
    R_0^{(p)} &\equiv \sum_{k=0}^{L-1} (k + 1) \frac{\Gamma(1 + pb)}{\Gamma(k+2 + pb)} \left(\frac{p}{2\beta_0}\right)^k f_k^{(p)} ,
    \label{eq:R0}
\end{align}
with the sum in the strength $R_0^{(p)}$ stopping at the last known order.
The expression for $R_0^{(p)}$ in \cref{eq:R0} is the same as derived by Komijani~\cite{Komijani:2017vep} from asymptotic (\emph{i.e.}, $l\gg L$) considerations. 
The new information of the derivation of this section~\cite{Kronfeld:2023jab}, beyond \cref{eq:asymp-rl}~\cite{Beneke:1998ui} with 
strength from \cref{eq:R0}~\cite{Komijani:2017vep}, is that \cref{eq:approx-rl} is a systematic approximation valid at any order.

\begin{figure}[b]
    \centering
    \includegraphics[width=0.75\textwidth]{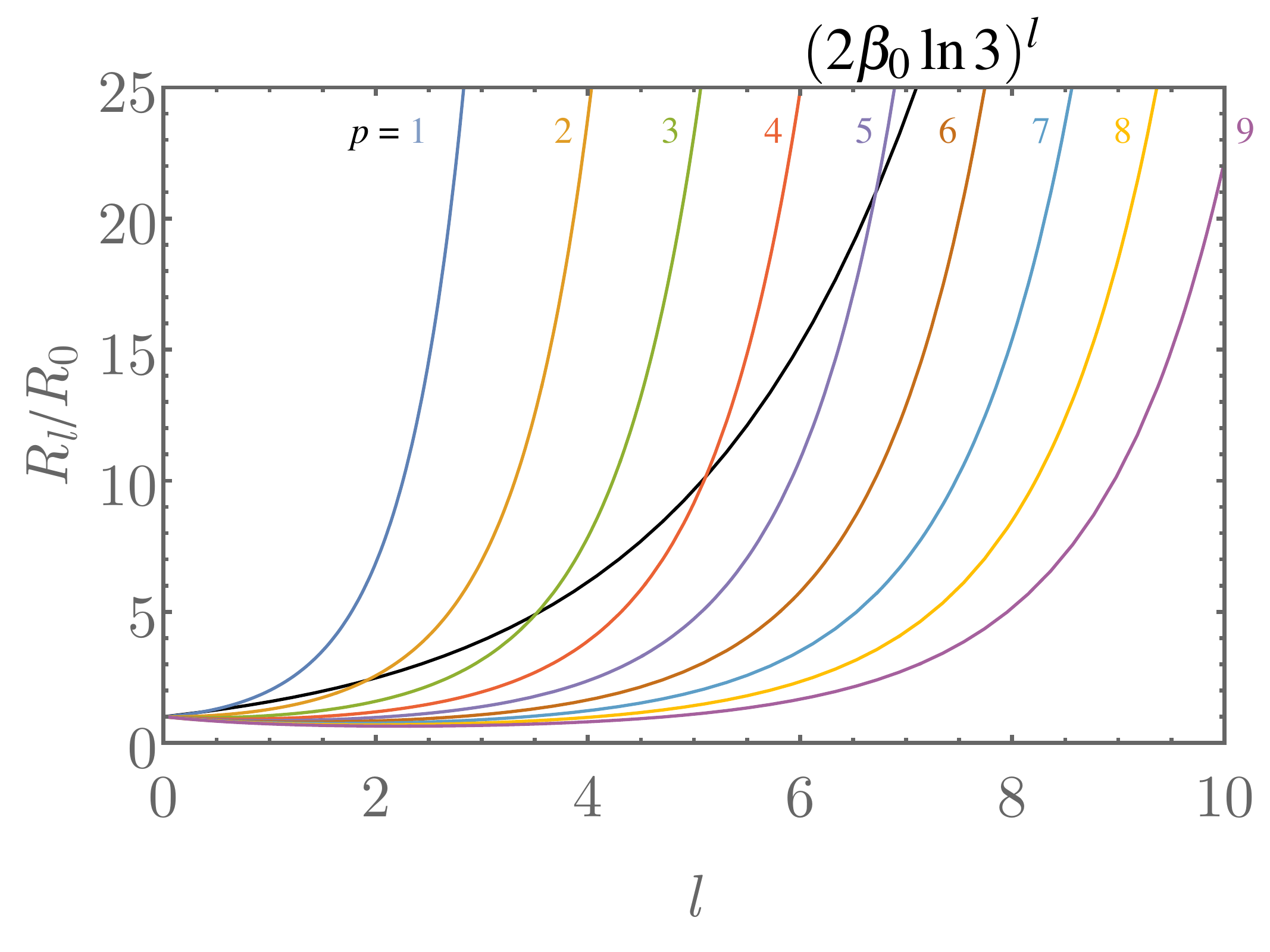}
    \caption[fig:growth]{Comparison of factorially growing terms in the perturbative series coefficients with a typical (if modest) 
        ``large'' logarithm.
        It seems at least as important to sum factorials (cf.\ \cref{sec:Borel}) as logs.
        $\beta_0$ and $b$ are taken for $n_f=3$.}
    \label{fig:growth}
\end{figure}
It is customary in perturbative QCD to sum large logarithms.
In simple problems of the form of \cref{eq:effcharge}, renormalization logarithms are summed by setting $\mu\propto Q$.
\Cref{fig:growth} compares the contributions to coefficients from a generic ``large log'' $(2\beta_0\ln3)^l$ with the factorial 
growth that \emph{also} stems from renormalization constraints.
The shapes of these curves suggests that is more important to sum the factorial growth than the logs.
Of course, both should be summed to the extent possible.

\section{Perturbative series reinterpreted}
\label{sec:PT}

\subsection{Better approximation}
\label{sec:better}

\Cref{eq:approx-rl,eq:R0} suggest a new approximation for the perturbative series%
\footnote{\Cref{eq:fk,eq:rl} hold for a scheme for \as\ in which the beta function takes a specific simple form, known as the 
``geometric scheme'' \cite{Brown:1992pk, Komijani:2017vep, Brambilla:2017hcq}.
In general \as\ schemes, one switches to the geometric scheme to obtain \cref{eq:rl} and switches back for \cref{eq:new}; for 
details, see ref.~\cite{Kronfeld:2023jab}.}
\begin{equation}
    R(Q) = \sum_{l=0}^{L-1} r_l(\mu/Q) \as(\mu)^{l+1} + \sum_{l=L}^\infty R^{(p)}_l(\mu/Q) \as(\mu)^{l+1} ,
\end{equation}
using the first $L$ $r_l$ from the literature and then their leading factorial growth for the rest of the series.
For $l\ge L$, \cref{eq:approx-rl} is certainly a better approximation than the standard truncation of setting unknown $r_l$ to~$0$.
It is convenient to set the lower limit of the second sum to~$l=0$,
\begin{equation}
    R(Q) = \sum_{l=0}^{L-1} \left[r_l(\mu/Q) - R^{(p)}_l(\mu/Q)\right]\as(\mu)^{l+1} +
        \sum_{l=0}^\infty R^{(p)}_l(\mu/Q) \as(\mu)^{l+1} .
    \label{eq:new}
\end{equation}
Reference~\cite{Brambilla:2017hcq} called this formula (as applied to the quark mass) ``minimal renormalon subtraction'' (MRS), 
because  the first sum is known from other work as the ``renormalon subtracted'' (RS) series~\cite{Pineda:2001zq}.
The second sum in \cref{eq:new} does not converge because of the factorially growing terms.
As with \cref{eq:tony}, the Borel summation procedure is applied, albeit with a twist to be discussed in \cref{sec:Borel}.

For the series relating the heavy-quark pole mass to its \MSbar\ mass, \cref{eq:new} was proposed already in 
ref.~\cite{Brambilla:2017hcq} and then applied to a precise determination of all quark masses (except top)~\cite{FermilabLattice:2018est}.
In ref.~\cite{Brambilla:2017hcq} we thought that \cref{eq:approx-rl} held only asymptotically~\cite{Komijani:2017vep}, so we introduced as the ``MRS prescription'' the subtraction and addition of the asymptotically high-order terms (even at intermediate orders) and then truncating the first series:
\begin{align}
    \sum_l r_l \as^{l+1} &\to \sum_l [r_l - R^{(p)}_l]\as^{l+1} + \sum_l R^{(p)}_l(\mu/Q) \as(\mu)^{l+1} \nonumber \\
        &\to \sum_{l=0}^{L-1} [r_l - R^{(p)}_l]\as^{l+1} + \sum_{l=0}^\infty R^{(p)}_l(\mu/Q) \as(\mu)^{l+1}.
\end{align}
This logic of ref.~\cite{Kronfeld:2023jab} is stronger than that of ref.~\cite{Brambilla:2017hcq}, simply because it is based on a 
systematic approximation at every order, \cref{eq:approx-rl}, rather than a prescription abstracted from asymptotically high orders.

\subsection{Borel summation}
\label{sec:Borel}

For understanding how the two the two sums in \cref{eq:new} complement each other, it is useful to write
\begin{align}
    R_\text{RS}(Q,\mu) &\equiv \sum_{l=0}^{L-1} \left[r_l(\mu/Q) - R^{(p)}_l(\mu/Q)\right]\as(\mu)^{l+1} ,
    \label{eq:RS} \\
    R_\text{B}(Q,\mu)  &\equiv \sum_{l=0}^\infty R^{(p)}_l(\mu/Q) \as(\mu)^{l+1} ,
    \label{eq:B}
\end{align}
so that
\begin{equation}
    R(Q) = R_\text{RS}(Q,\mu) + R_\text{B}(Q,\mu) .
\end{equation}
The subtracted series $R_\text{RS}(Q,\mu)$ and the Borel sum $R_\text{B}(Q,\mu)$ depend on the renormalization scale~$\mu$ but the 
sum does so only because of truncation; see \cref{sec:plots}.

The sum on the right-hand side of \cref{eq:B} is formal, so $R_\text{B}(Q,\mu)$ still requires a definition that can be evaluated.
A simple version of Borel summation is to express $\Gamma(l+1 + pb)$ via the integral representation $\Gamma(l+1+pb)=\int_0^\infty 
t^{l+pb}\e^{-t}\dd{t}$, carry out the sum, and then integrate over $t$
\begin{equation}
    R_\text{B}^{(p)} = \frac{R_0^{(p)} \as}{\Gamma(1 + pb)} \int_0^\infty \sum_{l=0}^\infty
        \left(\frac{2\beta_0\as t}{p}\right)^l t^{pb}\e^{-t}\dd{t}
        = \frac{R_0^{(p)} \as}{\Gamma(1 + pb)} \int_0^\infty \frac{t^{pb}\e^{-t}}{1-2\beta_0\as t/p}\dd{t};
\end{equation}
see the Appendix of ref.~\cite{Kronfeld:2023jab} for details.
There is a simple pole along the integration contour, but it suffices to take the principal part and define
\begin{equation}
    R_\text{B}^{(p)} = R_0^{(p)} \frac{p}{2\beta_0} \mathcal{J}(pb,p/2\beta_0\as) ,
    \label{eq:R-Borel-sum}
\end{equation}
where $\mathcal{J}$ is an analytic function with a rapidly convergent power series in the second argument,
$p/2\beta_0\as$~\cite{Brambilla:2017hcq}.

Beyond the principal value, one could deform the contour just above or below the real axis, obtaining an additional contribution
\begin{equation}
        - R_0^{(p)} \e^{\pm\iunit pb\pi} \frac{p^{1+pb}}{2^{1+pb}\beta_0} \Gamma(-pb)
            \left[\frac{\e^{-1/2\beta_0\as}}{(\beta_0\as)^b} \right]^p .
\end{equation}
The quantity in brackets is nothing but $\Lambda/Q$, so it can and should be absorbed into the power correction in the final 
prescription for factorially summed perturbation theory:
\begin{equation}
    \mathcal{R}(Q) = r_{-1} + R_\text{RS}(Q,\mu) + R_\text{B}(Q,\mu) + C_p \frac{\Lambda^p}{Q^p}.
    \label{eq:final}
\end{equation}
In practice, an \as\ determination consists of fitting data for $\mathcal{R}$ to the expressions on the right-hand side of \cref{eq:final} with fit parameters $\Lambda_{\MSbar}$ and~$C_p$.

\section{Lot of plots}
\label{sec:plots}

Let us now see how the factorial summation (or minimal renormalon subtraction~\cite{Komijani:2017vep, Brambilla:2017hcq}) fares in 
three examples for which $L=4$ orders of perturbation theory are available: the static energy, the pole mass, and the polarized 
Bjorken sum rule.
The results presented below rely on calculations of the coefficient in the perturbation series (cited in turn below) and the QCD 
beta-function~\cite{Gross:1973id, Politzer:1973fx, Jones:1974mm, Caswell:1974gg, Tarasov:1980au, Larin:1993tp, vanRitbergen:1997va, 
Czakon:2004bu, Zoller:2016sgq}; the five-loop coefficient is available~\cite{Baikov:2016tgj, Herzog:2017ohr, Luthe:2017ttc} but not 
needed here.

\subsection{Static energy}
\label{sec:E0}

The ``static energy'', denoted here $E_0(r)$, is the gluonic energy stored between a static quark and a static antiquark, separated by distance~$r$.
It is the energy of the lowest-lying state in a correlation function corresponding to a rectangular Wilson loop with spatial side~$r$.
This energy is often called the ``static potential'', but the potential energy is only part of the static energy.
Indeed, the potential on its own has infrared divergences~\cite{Appelquist:1977es}, which are canceled by a chromoelectric dipole 
contribution~\cite{Brambilla:1999qa, Brambilla:1999xf}---obviously the total static energy (from the Wilson loop) is infrared safe.

In potential nonrelativistic QCD, the static energy is given by a perturbation series and a power term:
\begin{equation}
    E_0(r) = - \frac{C_F}{r} \sum_{l=0} v_l(\mu r) \as(\mu)^{l+1} + \Lambda_0 .
    \label{eq:E0}
\end{equation}
The $r$-independent quantity $\Lambda_0$ is of order~\LQCD.
In the notation of \cref{eq:effcharge}, $\mathcal{R}(1/r)=-rE_0(r)/C_F$ is the dimensionless quantity at hand.
(The coefficients are denoted $v_l$ because the separation distance~$r$ appears as $1/r$ in place of $Q$ in the 
equations of \cref{sec:intro,sec:fgb,sec:fgrg,sec:PT}.)

In practice, perturbation theory is carried out first in momentum space
\begin{equation}
    \tilde{R}(q) = \sum_{l=0} a_l(\mu/q) \as(\mu)^{l+1} ,
\end{equation}
and the Fourier transform (of the $(\ln\mu/q)^l$ term in $a_l(\mu/q)$) generates the $p=1$ factorial growth indicated by \cref{eq:E0}.
The growth of the $a_l$, whatever it is, is characterized by $p>1$.

The series $\mathcal{F}(1/r)$ from \cref{eq:FQ} (dropping the superscript for brevity) is proportional to the ``static force'' 
$\mathfrak{F}\equiv-\dd E_0/\dd r$, which is thought to have no power corrections until those caused by QCD instantons, \emph{i.e.}, $p=9$ (for $n_f=3$).
Thus, this problem automatically provides series with $p=1$, $p>1$, and $p\ge9$, allowing study of very different behavior among intimately related sets of perturbative coefficients.

The perturbative coefficients have been calculated through order $\as^4$~\cite{Fischler:1977yf, Billoire:1979ih, Schroder:1998vy, Kniehl:2001ju, Smirnov:2008pn, Anzai:2009tm, Smirnov:2009fh, Lee:2016cgz, Brambilla:2006wp}.
\begin{table}[b]
    \centering
    \sisetup{table-format = 3.6, table-alignment-mode = format}
    \begin{tabular}{|c|SS|SS|SS|}
    \hline
    & \multicolumn{2}{c|}{\MSbar} & \multicolumn{2}{c|}{geometric} & \multicolumn{2}{c|}{$\an{2}$ \cite{Kronfeld:2023jab}} \\
    $l$ & {$a_l(1)$} & {$f_l(1)$} & {$a_l(1)$} & {$f_l(1)$} & {$a_l(1)$} & {$f_l(1)$} \\
    \hline
    $0$ &   1.       &  1.        &   1.       &  1.        &   1.       &  1.        \\
    $1$ &   0.557042 & -0.048552  &   0.557042 & -0.048552  &   0.557042 & -0.048552  \\
    $2$ &   1.70218  &  0.687291  &   1.83497  &  0.820079  &   1.83497  &  0.820079  \\
    $3$ &   2.43687  &  0.323257  &   2.83268  &  0.558242  &   3.01389  &  0.739452  \\
    \hline
    \end{tabular}
    \caption[tab:coeff-vf]{Perturbation series coefficients with $s=1$ for $\tilde{R}(q)$ ($a_l$) and $F(r)$ ($f_l$).
        From ref.~\cite{Kronfeld:2023jab}.}
    \label{tab:coeff-vf}
    \vspace*{2em}
    \begin{tabular}{|c|SS|SS|SS|}
    \hline
    & \multicolumn{2}{c|}{\MSbar} & \multicolumn{2}{c|}{geometric} & \multicolumn{2}{c|}{$\an{2}$ \cite{Kronfeld:2023jab}} \\
    $l$ & {$v_l(1)$} & {$v_l(1)-V_l(1)$} & {$v_l(1)$} & {$v_l(1)-V_l(1)$} & {$v_l(1)$} & {\!$v_l(1)-V_l(1)$} \\
    \hline
    $0$ &   1.       &      0.206061     &   1.       &      0.182531     &   1.       &      0.177584     \\
    $1$ &   1.38384  &     -0.202668     &   1.38384  &     -0.249689     &   1.38384  &     -0.259574     \\
    $2$ &   5.46228  &      0.019479     &   5.59507  &     -0.009046     &   5.59507  &     -0.042959     \\
    $3$ &  26.6880   &      0.219262     &  27.3034   &      0.050179     &  27.4846   &      0.066468     \\
    \hline
    \end{tabular}
    \caption[tab:coeff]{Perturbation series coefficients with $s=1$ for $R(r)$ and $R_\text{RS}$ (with $V_l$ derived from $v_l$ as
        $R_l$ from $r_l$ in \cref{sec:fgrg}).
        From ref.~\cite{Kronfeld:2023jab}.}
    \label{tab:coeff}
\end{table}
\Cref{tab:coeff-vf} shows these first four coefficients for the momentum-space quantity $a_l$ and the static force $f_l$, with 
$\mu=q$ and $1/r$, respectively.
The three schemes for \as\ are defined in ref.~\cite{Kronfeld:2023jab}.%
\footnote{The ``geometric'' scheme is the one in which the algebra is simplest, leading to \cref{eq:approx-rl,eq:R0,eq:new}.
The coupling $\an{2}$ is designed to lack a Landau pole (from which the \MSbar\ and the geometric scheme suffer).}
The qualitative features of the coefficients are the same in all three schemes.
Whether or not the $a_l$ are growing factorially cannot be divined from \cref{tab:coeff-vf}.
The $f_l$ do not appear to be growing in any dramatic way.

\Cref{tab:coeff} shows the first four coefficients of the static energy $v_l$ together with the subtracted coefficients, $v_l-V_l$, in $R_\text{RS}$.
The $v_l$ are obviously growing, and the amazing cancellation in $v_l-V_l$ argues that the growth is the $p=1$ factorial growth derived above.

\begin{figure}
    \centering
    \includegraphics[width=\figwidth,trim={0 50 0 0},clip]{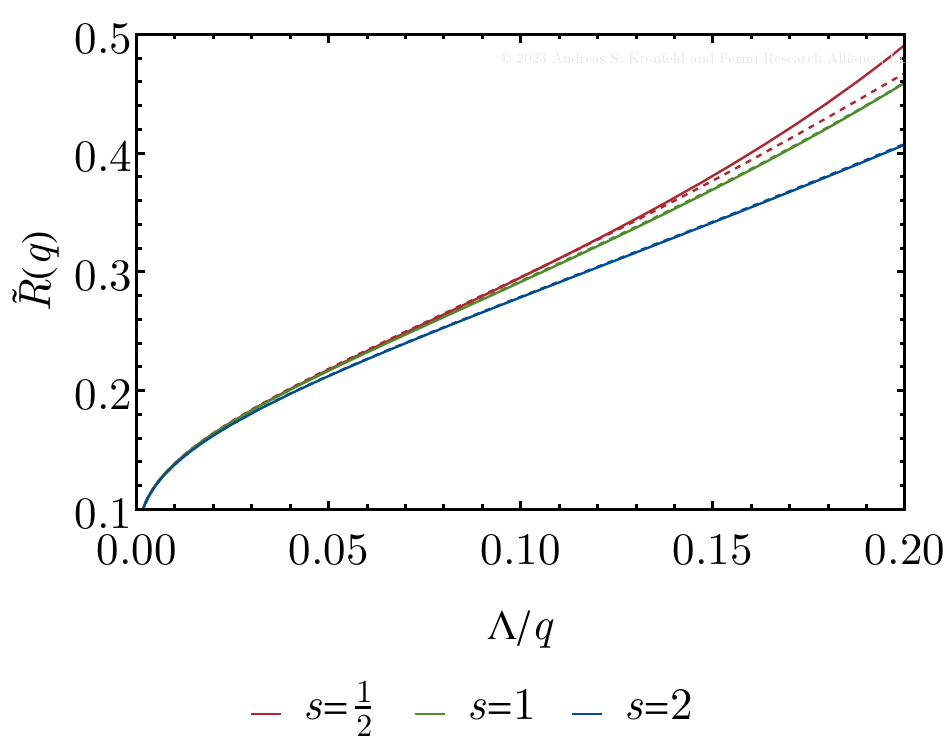} \hfill
    \includegraphics[width=\figwidth,trim={0 50 0 0},clip]{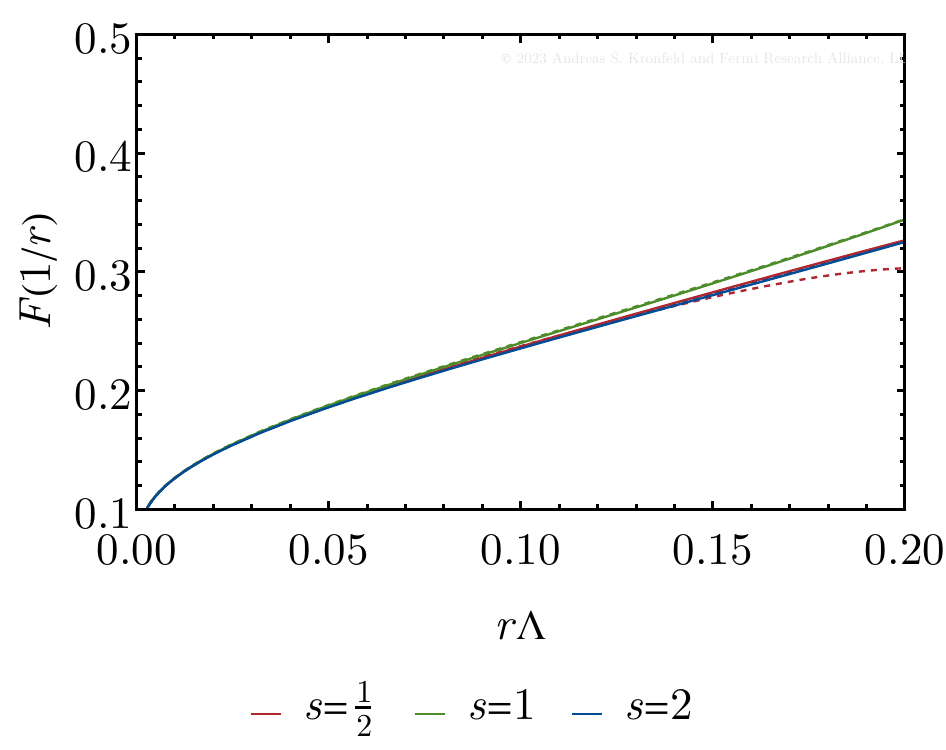} \\[6pt]
    \includegraphics[width=\figwidth,trim={0 50 0 0},clip]{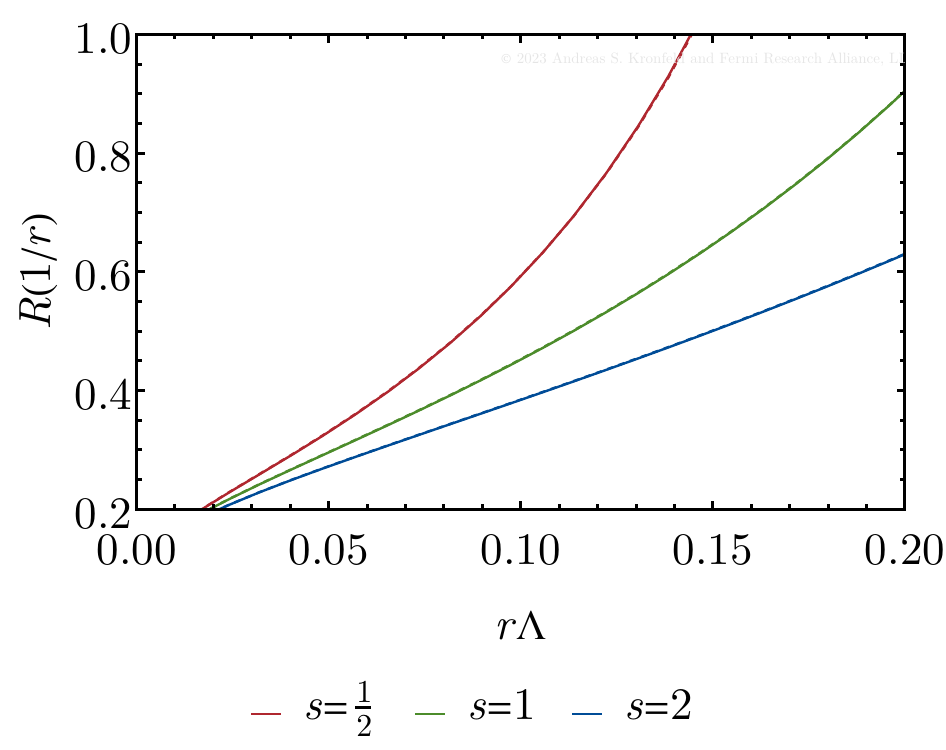} \hfill
    \includegraphics[width=\figwidth,trim={0 50 0 0},clip]{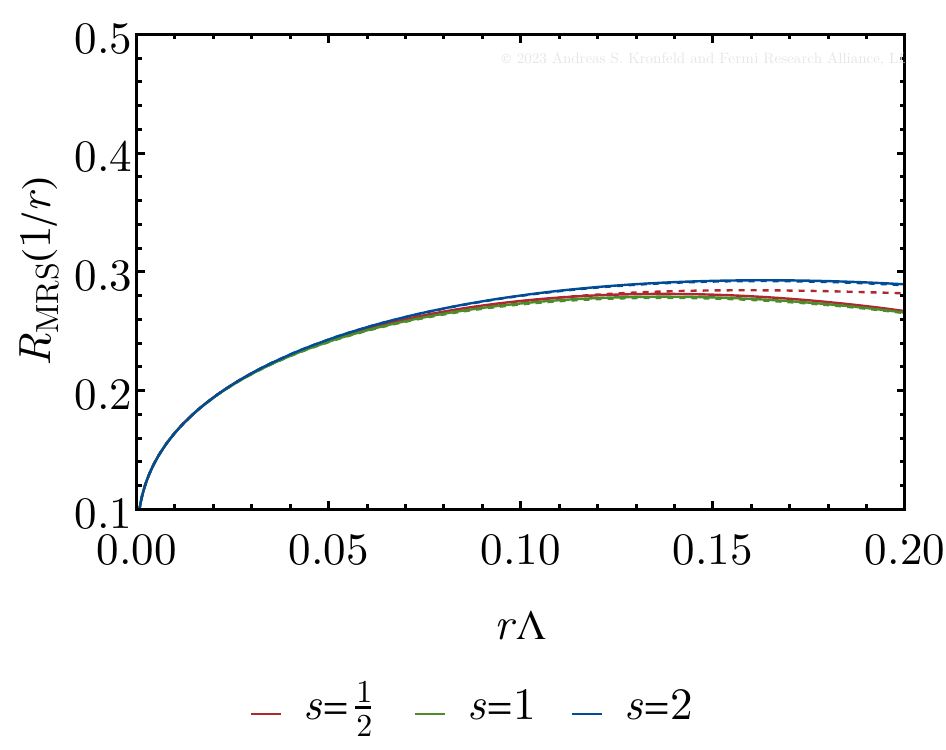}
    \caption[fig:PT]{Scale variation in the $\an{2}$ scheme.
        Top: $\tilde{R}(q)$ and $F(1/r)$; neither suffers the $p=1$ growth.
        Bottom: $R(1/r)$ (with $p=1$ growth) and $R_\text{MRS}(1/r)$ (after MRS).
        Red, green, and blue curves correspond to $s=\half$, $s=1$, and $s=2$, respectively.
        Solid (dashed) curves correspond to a running (fixed) \as\ in the ultrasoft $\ln\as$.
        Note that the vertical scale for $R(1/r)$ is twice that of the other three plots.
        From ref.~\cite{Kronfeld:2023jab}.}
    \label{fig:PT}
\end{figure}
\Cref{fig:PT} shows how well perturbation theory converges in the $\an{2}$ scheme for $\tilde{R}(q)$, $F(1/r)$, fixed-order
$R(1/r)$, and $R_\text{MRS}(1/r)\equiv R_\text{RS}(1/r)+R_\text{B}(1/r)$.
The curves show the quantities as a function of $\Lambda r$ (or $\Lambda/q$), obtained by expressing \as\ as a function of
$\ln\Lambda r$ (or $\ln\Lambda/q$), adjusted so that $\Lambda=\Lambda_{\MSbar}$.
Plotted this way, the high-energy/short-distance regime is squeezed to the left, making it easy to see region in which perturbation 
theory fails.
The different colors show $\mu=s/r$ (or $\mu=sq$ for $\tilde{R}(q)$) for $s=\half$, $1$, and $2$---the degree of agreement for
varying $s$ tests whether the truncation at four terms is successful.
The verdicts are, respectively, pretty good, great, horrible, and great again.

The cancellation seen in \cref{tab:coeff} is most spectacular when choosing $\mu=1/r$ but still impressive otherwise.
\begin{figure}
    \centering
    \includegraphics[width=\figwidth,trim={0 50 0 0},clip]{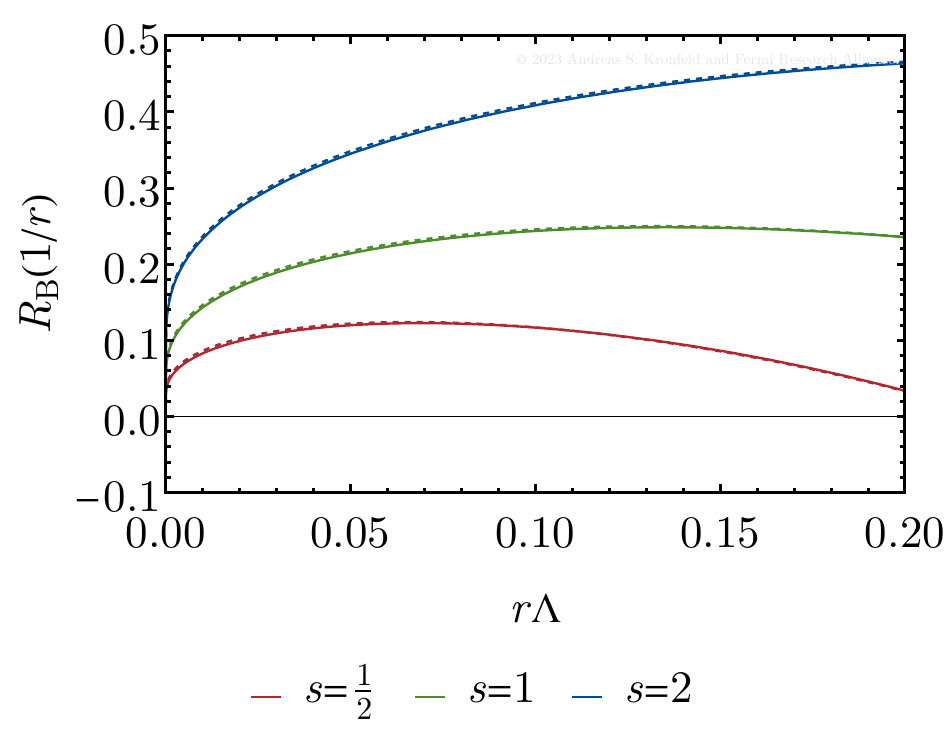} \hfill
    \includegraphics[width=\figwidth,trim={0 50 0 0},clip]{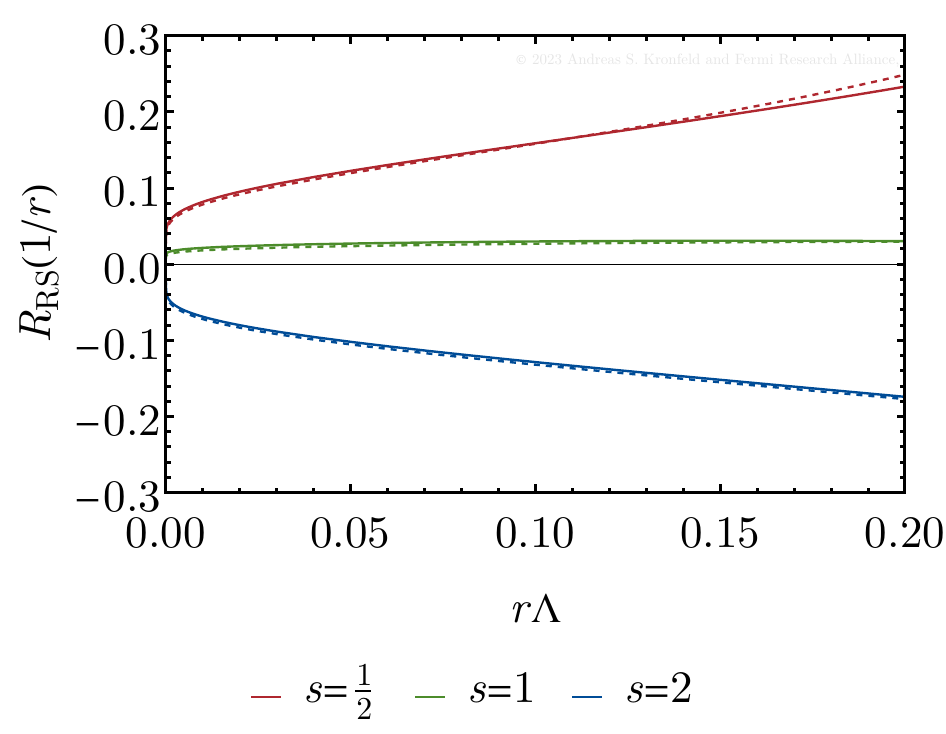}
    \caption[fig:cancel]{Scale variation in the \an{2} scheme of the Borel sum $R_\text{B}(1/r)$~(left) and the
        $L=4$-sub\-tracted series $R_\text{RS}(1/r)$~(right).
        Curve and color code as in \cref{fig:PT}.
        From ref.~\cite{Kronfeld:2023jab}.}
    \label{fig:cancel}
\end{figure}
\Cref{fig:cancel} shoes the scale variation for $s=\half$, $1$, and $2$ for the Borel sum $R_\text{B}(1/r)$ and the $L=4$-subtracted series $R_\text{RS}(1/r)$.
While both vary with~$\mu$ (\emph{i.e.}, with~$s$), the variation cancels when adding to get the lower-right panel of \cref{fig:PT}.
To get a feel for the cancellation of the factorial growth for $s=\half$ and $2$, compare $R(1/r)$ in the lower-left panel of \cref{fig:PT} the curves for $R_\text{RS}(1/r)$ in \cref{fig:cancel}: at $r\Lambda=0.1$, $R(10\Lambda)\approx0.4$--$0.6$ while $R_\text{RS}(10\Lambda)\approx\pm0.1$ (and an order of magnitude small still for $s=1$).

An important feature of $R_\text{MRS}(1/r)$ is that in the region where the constant~$\Lambda_0$ in \cref{eq:E0} is 
significant---in $\mathcal{R}$ it turns into $-r\Lambda_0/C_F$---it is straightforward for a fit to distinguish the power term from 
the perturbation series.
Distinguishing linear in $r$ from the lower-left of \cref{fig:PT} would be hopeless.
The TUMQCD Collaboration~\cite{Leino:2025pvl} is fitting lattice-QCD data for the static energy, which have been 
obtained~\cite{Brambilla:2022het} on ensembles from the MILC Collaboration with $2+1+1$ flavors of sea quark~\cite{MILC:2012znn, Bazavov:2017lyh}.

\subsection{Quark pole mass}
\label{sec:pole}

Here the series at hand is the relation between the renormalized \MSbar\ mass and the pole mass.
The analog to \cref{eq:effcharge} is the relation between a heavy-light hadron mass $\mathcal{M}$ and the quark mass
\begin{equation}
    \mathcal{M} = M + \bar{\Lambda} + \frac{\mu_\pi^2}{2M} + \cdots,
    \label{eq:hqet}
\end{equation}
where the pole mass $M$ is related to the \MSbar\ mass $\bar{m}=m_{\MSbar}(\mu)|_{\mu=\bar{m}}$ by
\begin{equation}
    M = \bar{m} \left[1 + \sum_{l=0} r_l \as(\bar{m})^{l+1} \right].
    \label{eq:pole}
\end{equation}
The coefficients $r_l$ grow as in \cref{eq:approx-rl} with $p=1$, related to the energy of gluons and light quarks,~$\bar{\Lambda}$.
\Cref{eq:hqet} is not quite the same as \cref{eq:effcharge}, because the higher powers are suppressed not by a physical energy $Q$ 
but by an object expressed as an ambiguous series, \cref{eq:pole}.
\begin{figure}[b]
    \includegraphics[width=\figwidth,trim={0 50 0 0},clip]{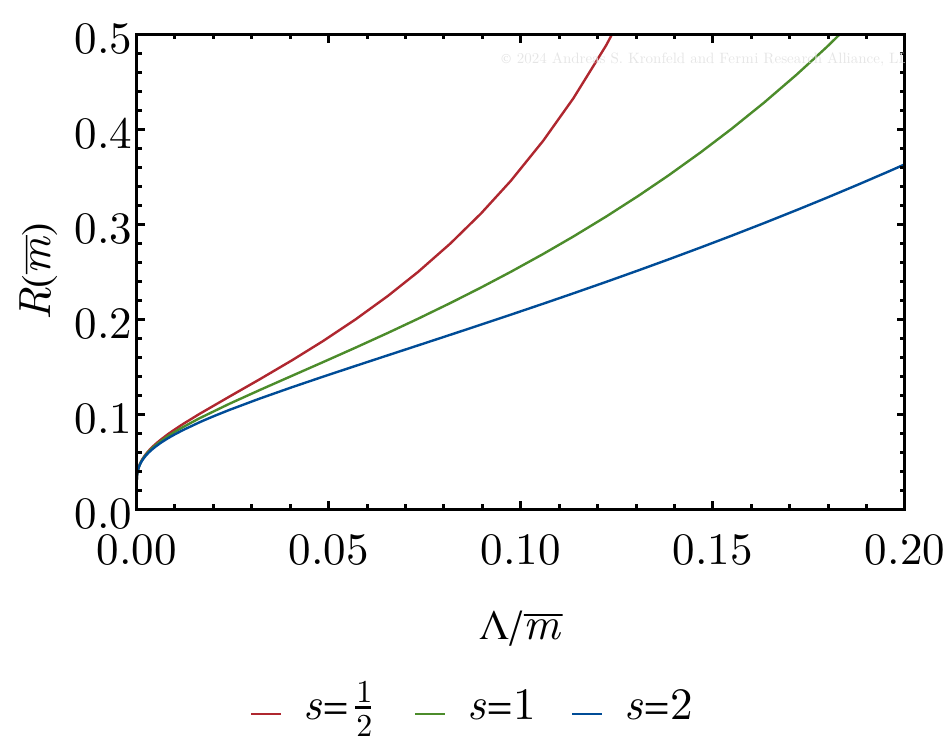}\hfill
    \includegraphics[width=\figwidth,trim={0 50 0 0},clip]{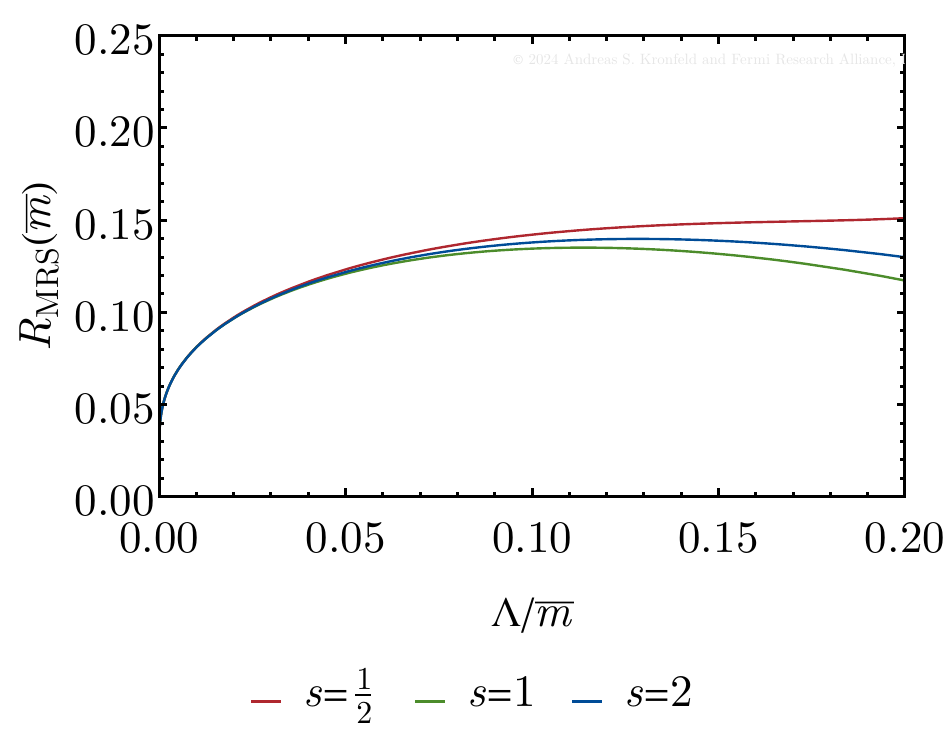}
    \caption[fig:good]{Left: fixed-order perturbation theory for $R(\bar{m})\equiv M/\bar{m}-1$ vs.\ $\Lambda/\bar{m}$
        for $\mu=s\bar{m}$; right: factorially summed $R_\text{MRS}(\bar{m})$.
        In both, the \MSbar\ scheme is used for \as, and $s\in\{\half,1,2\}$ with the same color code as in \cref{{fig:PT}}.
        From ref.~\cite{Kronfeld:2024qao}.}
    \label{fig:pole}
\end{figure}
Fixed-order ($L=4$) and MRS definitions of the series are shown in \cref{fig:pole}.%
\footnote{The two-, three-, and four-loop coefficients have been computed in refs.~\cite{Gray:1990yh, Chetyrkin:1999ys, Chetyrkin:1999qi, Melnikov:2000qh, Marquard:2015qpa, Marquard:2016dcn}.}
The fixed-order estimate of $R(\bar{m})\equiv M/\bar{m}-1$ is a disaster, but after summing factorials the variation 
with the choice of $s$ that is almost as mild as the static force or summed static energy in \cref{fig:PT}.
The slightly larger variation is connected to the kinetic energy term $\mu_\pi^2/2M$ is \cref{eq:hqet}.

The MRS mass was used to determine the charm and bottom quark masses with sub percent precision from lattice-QCD calculations of 
heavy-light meson masses~\cite{FermilabLattice:2018est}.

Because a quarkonium mass to leading accuracy is given by $\langle2M+E_0\rangle$ (where the angle brackets indicated expectation
value in a quarkonium state), it would be satisfying if the factorial strengths of the two were the same.
This is a nontrivial test, because the coefficients of the two series look completely different.
Explicit evaluation of the strengths yields $2R_0(s=\{\half,1,2\})=\{0.512564,1.06359,2.0329\}$ and 
$C_FV_0(s=\{\half,1,2\})=\{0.544191, 1.09655, 2.07881\}$ (here in the $\an{2}$ scheme).

\subsection{Bjorken sum rule}
\label{sec:Bj}

Another quantity with $L=4$ terms available in the literature is the nonsinglet, polarized Bjorken sum rule~\cite{Bjorken:1966jh, 
Bjorken:1969mm} in deep inelastic scattering.
At large momentum transfer~$Q$, perturbative QCD~\cite{Kodaira:1978sh,Kodaira:1979ib} finds
\begin{equation}
    \Gamma_1^{(p-n)}(Q) = \frac{g_A}{6}\left(1 - \frac{3}{4\pi}C_F \sum_{l=0}r_l(\mu/Q)\as(Q)^{l+1}\right) +
        \sum_{\text{even}~p} C_p \frac{\Lambda_p}{Q^p},
    \label{eq:Bj}
\end{equation}
where $g_A$ is the nucleon axial charge.
$\Gamma_1$ is an integral over all Bjorken~$x$ of the polarized structures function~$g_1(x)$, and the superscript implies the 
difference between proton and neutron measurements; see \emph{e.g.}\ ref.~\cite{Blumlein:2012bf} for details.
The perturbation series is augmented with higher-twist terms with $p=2$, $4$, $6$, \ldots.

The coefficients $r_l$ have been computed through order $\as^4$~\cite{Gorishnii:1983gs, Larin:1990zw, Baikov:2010je, Chetyrkin:2022fqk}.
Using them (and the beta function coefficients), the standard truncated and factorially improved approximations to 
\begin{align}
    \mathcal{R}(Q) &\equiv \frac{4\pi}{3C_F}\left(1 - \frac{6}{g_A} \Gamma_1^{(p-n)}(Q) \right) \text{ and} \\
    R(Q) &\equiv \sum_{l=0}r_l(\mu/Q)\as(Q)^{l+1}
    \label{eq:BjR}
\end{align}
can be constructed.
\begin{figure}
    \includegraphics[width=\figwidth,trim={0 50 0 0},clip]{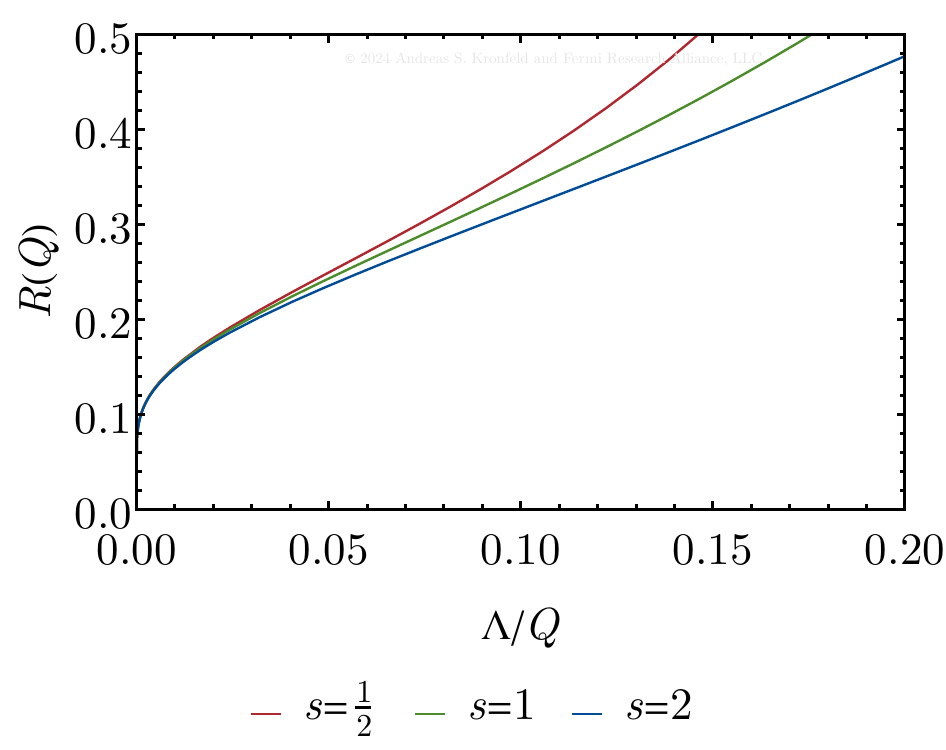}\hfill
    \includegraphics[width=\figwidth,trim={0 50 0 0},clip]{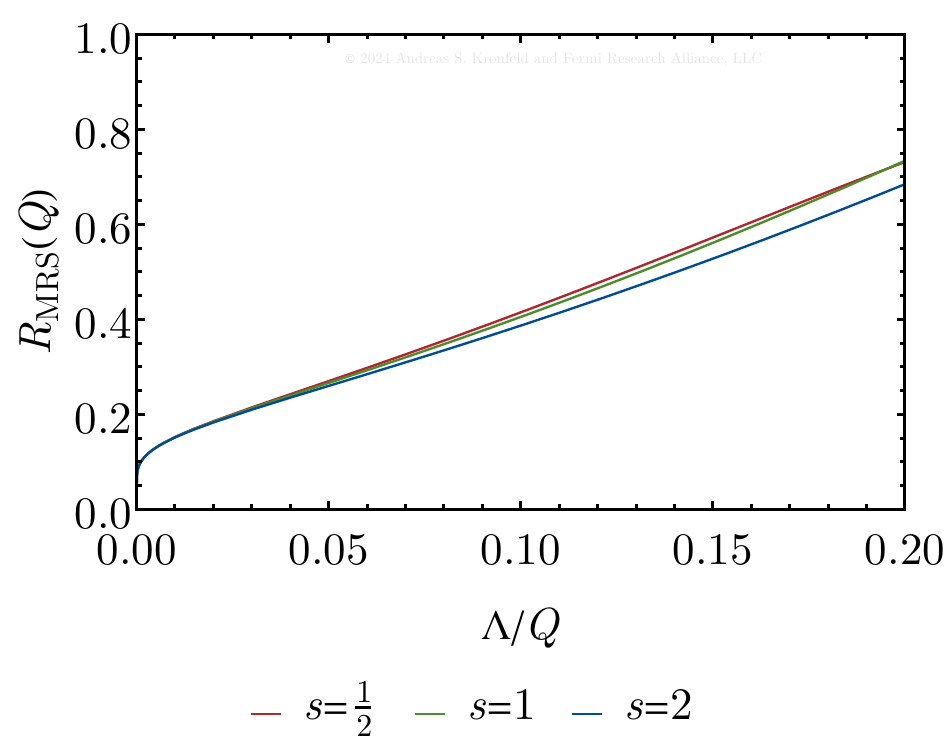} \\
    \includegraphics[width=\figwidth,trim={0 50 0 0},clip]{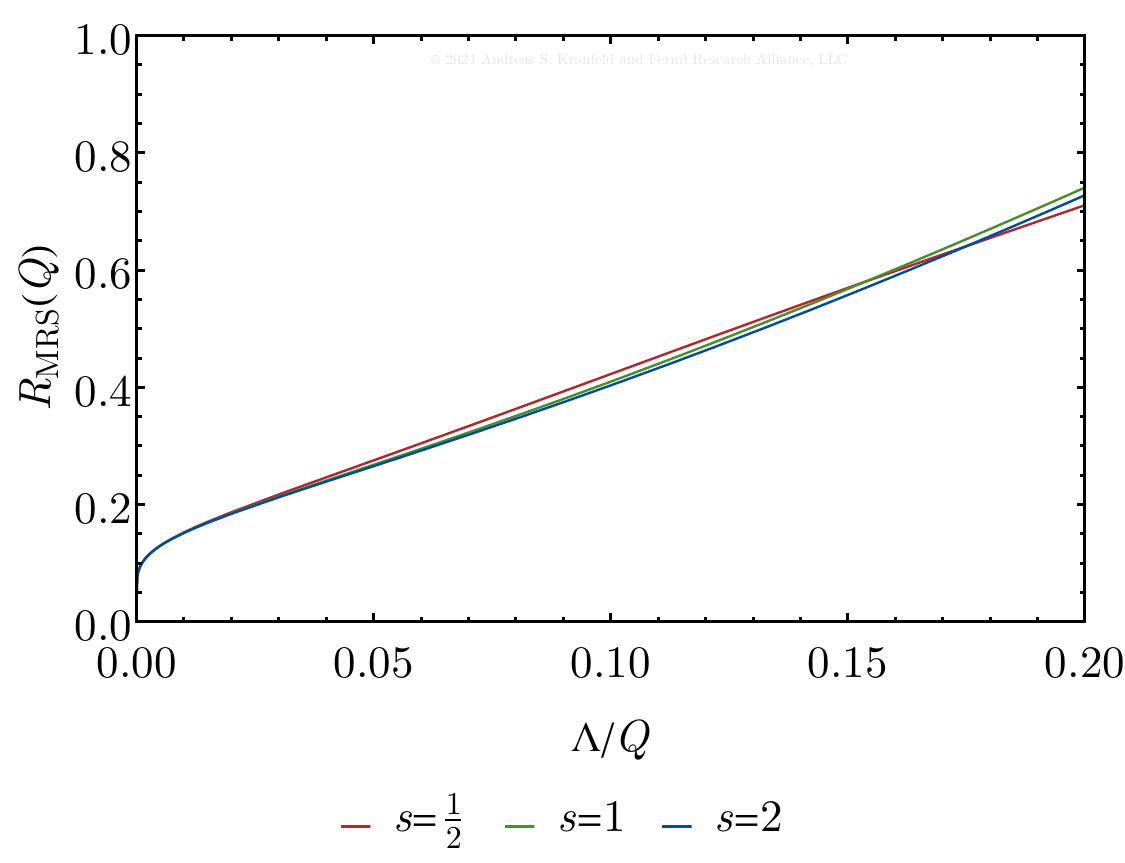}\hfill
    \includegraphics[width=\figwidth,trim={0 50 0 0},clip]{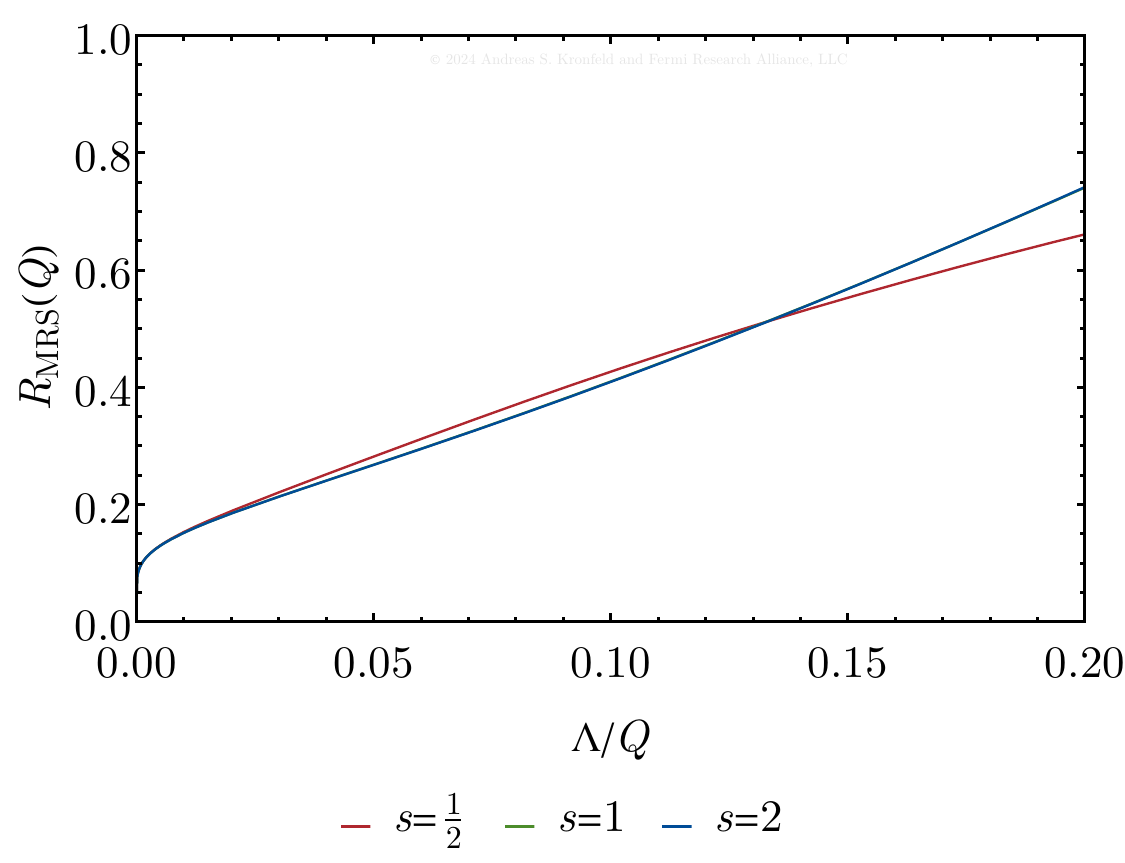}
    \caption[fig:Bj]{\emph{Preliminary!}
        Top    left:  fixed-order perturbation theory for $R(Q)$ in \cref{eq:BjR};
        top    right: MRS factorial sum for growth with $p=2$;
        bottom left:  MRS factorial sum for growth with $(p_1,p_2)=(2,4)$;
        bottom right: MRS factorial sum for growth with $(p_1,p_2,p_3)=(2,4,6)$.
        In all cases, $\mu=sQ$, and the curves are for $s\in\{\half,1,2\}$ with the same color code as in \cref{{fig:PT}}.}
    \label{fig:Bj}
\end{figure}
\Cref{fig:Bj} shows the unsubtracted series as well as the result of summing the factorials associated with $p=2$ only, $(p_1,p_2)=(2,4)$, and $(p_1,p_2,p_3)=(2,4,6)$.
Although the formulas have not been presented in these proceedings, it is straightforward to generalize 
\cref{eq:approx-rl,eq:R0,eq:R-Borel-sum} to the case with more than one power.
\Cref{fig:Bj} shows improving stability when the $p=2$ and $(p_1,p_2)=(2,4)$ factorials have been summed, but also removing the 
third power correction (\emph{i.e.}, the $(p_1,p_2,p_3)=(2,4,6)$ case) does not help much.
A possible reason is that the the powers are not pure integers: the $Q$ dependence of the higher-twist terms is modified by anomalous dimensions.
Fractional dimensions are not a problem, but once the dimensions develop $Q$ dependence, the treatment of more than one power correction becomes cumbersome~\cite{Kronfeld:2023jab}.

It has been proposed to determine $\as$ via \cref{eq:Bj} using data from a future electron-ion collider~\cite{Kutz:2024eaq}.
There are also published data at not-so-large~$Q$.
Unfortunately, many reported results for $g_1^{(p)}$ and $g_1^{(n)}$ are from data taken on a set of points in the $Q$-$x$ plane 
without constant $Q$, with perturbative running being used to adjust measurements from the measured to a fixed value of~$Q$.
Determining \as\ from such inputs would be a tautology.
Work is in progress on this front.

\section{Outlook}

Reference~\cite{Kronfeld:2023jab} (and these proceedings) consider the ``minimal renormalon subtraction'' 
(MRS)~\cite{Komijani:2017vep, Brambilla:2017hcq} procedure, finding the generalization to any sequence of power corrections, 
thereby summing the dominant and subdominant factorial growth of the perturbation series.
The theoretical keystone is simple and basic, following the consequences of physical quantities being independent of the 
artificial scale~$\mu$ in \MSbar\ renormalization.
As it it standard in perturbative QCD to sum logarithms, ref.~\cite{Kronfeld:2023jab} (and \cref{sec:fgrg,sec:PT}) provides 
formulas to sum factorials, which grow even faster.
The mild variation in \cref{fig:Bj} when the $\mu$ is varied of over the customary range suggests truncation uncertainties can now 
be controlled.
Note that based on current understanding ``subtraction'' is not involved, rather a summation of known contributions is explicitly 
carried out.

\acknowledgments

This work was carried out with the partial support of the Technical University of Munich – Institute for Advanced Study, funded by the German Excellence Initiative.
Fermilab is managed by Fermi Forward Discovery Group, LLC, acting under Contract No.~89243024CSC000002.

\bibliographystyle{JHEP-x}
\bibliography{j,x}

\end{document}